# Spin current generation from an epitaxial tungsten dioxide $WO_2$

Kohei Ueda[1,2,3,a)], Hayato Fujii[1], Takanori Kida[4], Masayuki Hagiwara[4], and Jobu Matsuno[1,2,3]

*[1]Department of Physics, Graduate School of Science, Osaka University, Osaka 560-0043, Japan*
*[2]Center for Spintronics Research Network, Graduate School of Engineering Science, Osaka University, Osaka 560-8531, Japan*
*[3]Spintronics Research Network Division, Institute for Open and Transdisciplinary Research Initiatives, Osaka University, Osaka, 565-0871, Japan*
*[4]Center for Advanced High Magnetic Field Science, Graduate School of Science, Osaka University, Osaka 560-0043, Japan*

## ABSTRACT

We report on efficient spin current generation at room temperature in rutile type $WO_2$ grown on $Al_2O_3(0001)$ substrate. The optimal $WO_2$ film has (010)-oriented monoclinically distorted rutile structure with metallic conductivity due to $5d^2$ electrons, as characterized by x-ray diffraction, electronic transport, and x-ray photoelectron spectroscopy. By conducting harmonic Hall measurement in $Ni_{81}Fe_{19}/WO_2$ bilayer, we estimate two symmetries of the spin-orbit torque (SOT), i.e., dampinglike (DL) and fieldlike ones to find that the former is larger than the latter. By comparison with the $Ni_{81}Fe_{19}/W$ control sample, the observed DL SOT efficiency $\xi_{DL}$ of $WO_2$ (+0.174) is about two thirds of that of W (−0.281) in magnitude, with a striking difference in their signs. The magnitude of the $\xi_{DL}$ of $WO_2$ exhibits comparable value to those of widely reported Pt and Ta, and Ir oxide $IrO_2$. The positive sign of the $\xi_{DL}$ of $WO_2$ can be explained by the preceding theoretical study based on the 4*d* oxides. These results highlight that the epitaxial $WO_2$ offers a great opportunity of rutile oxides with spintronic functionalities, leading to future spin-orbit torque-controlled devices.

[a)]Author to whom correspondence should be addressed: kueda@phys.sci.osaka-u.ac.jp

Recently, strong spin-orbit coupling (SOC) of 5*d* electrons has attracted interest for emerging spin related functionality. One of the promising functionalities is a generation of the spin current that can be converted by charge current via the bulk spin-Hall effect (SHE)[1] of non-magnet with strong SOC, enabling the development of the magnetic device application[2]. An efficient spin current generation via the SHE has been widely investigated for 5*d* transition metals (TMs) such as Pt[3–10], Ta[11–13], and W[14–16]. More recently, an efficient spin current generation is also found in 5*d* Ir oxides[17–23] possibly due to its unique electronic structure dominated only by 5*d* electrons; the 5*d* transition metal oxides (TMOs) contrast well with the 5*d* TMs of which the electronic structure is dominated by both 6*s* and 5*d* electrons. Much effort has been so far made in epitaxial $SrIrO_3$[17–20], epitaxial $IrO_2$[21], and amorphous $IrO_2$[22,23], highlighting 5*d* TMOs as a class of spintronic materials.

Despite the importance of 5*d* TMOs in terms of spintronics, only the Ir oxides with $Ir^{4+}$ valence state ($5d^5$) have been a subject of the spintronic study because of their high conductivity among TMOs and their high chemical stability[24–26]. The situation motivates us to explore conductive 5*d* oxides with different 5*d* electron numbers as efficient spin current generator; we then expect that we can find a better material in a variety of 5*d* TMOs as well as we can obtain deeper insight into the spin-current physics related to 5*d* electrons. Of particular interest is a binary tungsten dioxide $WO_2$ with $W^{4+}$ valence state ($5d^2$)[27–31] since the $WO_2$ has been known as a good conductor which provides a great opportunity to examine the spin current generation. In contrast to the $WO_2$, highest-oxidation phase $WO_3$ with $W^{6+}$ valence state[32–34] is $5d^0$ band insulator which is significantly more thermodynamically stable. It is thus essential to control the phase stability of the $WO_2$ film by fine-tuning oxidizing conditions[30,31]. Within this context, previous reports have demonstrated the epitaxially grown $WO_2$ with metallic character[30,31], allowing us to attack an open question whether or not 5*d* TMOs other than Ir oxides can be an efficient spin current generator.

In this article, we report on room-temperature spin current generation from the epitaxial $WO_2$ grown on $Al_2O_3$(0001). We first characterize the optimal $WO_2$ film by investigating the crystal structure, resistivity, and electronic structure. Second, the spin-orbit torque (SOT) with dampinglike (DL) and fieldlike (FL) symmetries is quantified by harmonic Hall measurement. The measurement reveals that sizable DL- and FL-SOTs in $Ni_{81}Fe_{19}/WO_2$ bilayer and control sample $Ni_{81}Fe_{19}$/W bilayer, while the DL SOT is much larger than the FL SOT in both bilayers. The observed DL SOT with positive sign in $WO_2$ is smaller than that with negative sign in W. The magnitude of the DL SOT is comparable to reported values found in Pt and $IrO_2$, suggesting an efficient spin current arising from the epitaxial $WO_2$. These findings highlight that the unique electronic structure of $WO_2$ dominated by strong SOC plays an important role in spin current physics.

Figure 1(a) displays a crystal structure of the stoichiometric $WO_2$, which has a monoclinically distorted-rutile structure[27]. The lattice vector represents a monoclinic ($M$) and a pseudorutile ($R$) unit cell[27,30]: $a_M \sim 2c_R$, $b_M \sim a_R$, and $c_M \sim (b_R^2 + c_R^2)^{1/2}$ , where $a_M = 0.556$ nm, $b_M = 0.490$ nm, $c_M = 0.556$ nm, and $\beta = 120.47°$. This monoclinic unit cell has pseudo-trigonal symmetry, namely, combination of $a_M \sim c_M$ and $\beta \sim 120°$, which is inherent to the rutile structure. This is shown in the simplified sketch for surface oxygen-oxygen distance, which helps to understand the crystallographic relationship between the $WO_2$ and the $Al_2O_3$ as discussed later. The $Al_2O_3$ is known to have the corundum structure, which has the hexagonal bravais lattice with the lattice constant of 0.476 nm[35,36]. While the gray area represents the $b$-axis-orientated $WO_2$ plane, its rotation of 30 degree corresponds to the red area on the (0001)-oriented $Al_2O_3$ plane.

We fabricate the $WO_x$ films grown on $Al_2O_3$ (0001) substrates by a reactive magnetron sputtering of a W target at the working pressure of 0.4 Pa with Ar and oxygen. The sputtering parameters are sputtering power ($P$) and partial oxygen pressure; the former and the latter is 125 and 150 W, and 10–30%, respectively. Substrate temperature was fixed at 550℃. While the crystalline quality of the $WO_x$ films was confirmed by x-ray diffraction (XRD) measurement, the thickness was confirmed by x-ray reflectivity (XRR) measurements; all of the examined films have thickness of 20–30 nm. Figure 1(b) shows the out-of-plane XRD pattern of an epitaxial $WO_2$ film grown at optimized growth conditions, 150 W and 15 %. We observe two peaks, (020) and (040), of the monoclinic $WO_2$ without any secondary phases. Figure 1(c) [bottom] is a magnified view around the $WO_2$ (020) and the $Al_2O_3$(0006) peaks. The $WO_2$ (020) peak gives the out-of-plane lattice constant of 0.488 nm, which is close to ~0.490 nm for the bulk form[29–31]. The Laue fringes around the (020) diffraction evidence a good crystallinity of the film, which is also supported by narrow full width at half maximum (FWHM) of the rocking curve around the $WO_2$(020) (~0.018 deg.). Since the crystallinity is degraded to 0.028 deg. at lower oxygen ratio of 10%, we consider that the film grown at 15% has a better quality. In Fig. 1(c) [top], the film grown at 150 W and 30% has a relatively small diffraction peak close to the $Al_2O_3$ (0006) peak. The diffraction matches with bulk $WO_3$(111) in pseudocubic notation[33,34]; the $WO_3$ typically forms a perovskite structure without A site cation[32–34]. The $WO_3$ (111) film is relaxed, judging from the wide FWHM of the rocking curve around $WO_3$ (111) (~1.1 deg.). Calculated lattice constant of 0.380 nm is slightly longer than the bulk lattice constant 0.376 nm[32–34], resulting from the oxygen vacancy; we roughly estimate $x$ to be 2.97 based on the relation between the bulk lattice constant and amount of the oxygen vacancy[32].

In order to understand the in-plane crystalline orientation of the $WO_2$ film, we measured azimuthal $\phi$ scan at the $WO_2$ (011) [top] and $Al_2O_3$ $(10\bar{1}4)$ [bottom] diffractions as shown in Fig. 1(d). The diffraction of $Al_2O_3$ $(10\bar{1}4)$ presents three-fold rotational symmetry with 120° step, consistent with the three-fold axis of the corundum structure[36]. The diffraction of the $WO_2$ (011) exhibits six peaks with 60° step and shifts in 30° with respect to the $Al_2O_3\,(10\bar{1}4)$ in accordance with the epitaxial relationship in Fig. 1(a); the hatched area of the $WO_2$ (011) is rotated in 30° relative to that of $Al_2O_3\,(10\bar{1}4)$. We ascribe the observed six peaks of the $WO_2$ (011) to the formation of three-type $WO_2$ domains stemming from the trigonal symmetry of the $Al_2O_3$ substrate; each domain gives rise to two peaks corresponding to two-fold symmetry of the monoclinic unit cell.

Electronic transport in our $WO_x$ films also provides an important information. In order to examine how the resistivity varies with sputtering conditions, we measured room-temperature resistivity by carrying out the van der Pauw technique on all of the films. Figure 2(a) shows the resistivity as a function of the oxygen ratio in cases of sputtering power $P$ = 125 and 150 W. The resistivity is ~0.82 mΩcm for $WO_2$ and ~7300 mΩcm for $WO_3$, corresponding to the lowest and highest values at $P$ = 150 W. Since the resistivity of the $WO_2$ records less than 1 mΩcm, we consider our $WO_2$ film as a conductor. Irrespective of $P$, the resistivity shows a moderate increase at lower oxygen ratio and a drastic increase at higher oxygen ratio. While the low oxygen ratio promotes $WO_{2-\delta}$, the high oxygen ratio does $WO_{3-\delta}$ or $WO_3$; the $\delta$ represents the degree of oxygen vacancy. The oxygen ratio associated with the lowest resistivity decreases with the $P$, naturally explained by the fact that lower $P$ corresponds to higher oxygen ratio. We also note that the higher $P$ gives rise to lower resistivity at around the oxygen ration of 15% while there is no discernible difference of the crystallinity as a function of $P$; the optimal $P$ is concluded to be 150 W in terms of resistivity. Thus, the relation between $P$ and oxygen ratio is crucial for controlling the $WO_x$ phase, as demonstrated in the previous study[31].

We conducted x-ray photoelectron spectroscopy (XPS) measurement to illustrate the effect of electron filling for $WO_2$ and $WO_3$. The XPS measurements were performed with an Al $K_\alpha$ radiation; the binding energy was calibrated with respect to the C 1s peak at 284.8 eV. Figure 2(b) shows the valence band spectra for the $WO_2$ [bottom] and for the $WO_3$ [top]. The electronic structure of the $5d^2$ $WO_2$ is dominated by two electrons in $t_{2g}$ bands near the Fermi energy ($E_F$). When the rutile type $WO_2$ is formed, the degeneracy of the $t_{2g}$ orbitals are lifted due to the metal-metal bonding in the edge-sharing $WO_6$ octahedra[29]; the one electron is accommodated to the most stable $\sigma$- bonding band whereas the other electron is accommodated to the π-bonding band and higher-energy bands. We actually observe the double peaks at around 2 and 0.8 eV near $E_F$, corresponding to the $\sigma$-bonding and the π-bonding bands, respectively[35–37]. This manifests that our $WO_2$ film is a $5d^2$ conductor with the rutile structure while the conducting nature is supported as well by black color of the film shown in the inset. On the other hand, the electronic structure of the $5d^0$ $WO_3$ has no electron in $t_{2g}$ bands near the $E_F$. The inset shows that the film is almost transparent and hence is close to insulator. The valence band spectrum indicates a tiny spectral weight of the W 5*d* electron near $E_F$ stemming from the oxygen vacancy, which is consistent with the XRD and the resistivity results. To summarize the film characterization, we fabricated epitaxial rutile-type $WO_2$ film with $5d^2$ electrons dominating its conducting nature; the film is suitable for measuring the spin current generation.

We focus on observation of the SOT arising from the $WO_2$, which is manifestation of the spin current generation[37]. We prepared bilayer consisting of epitaxial $WO_2$ and ferromagnetic $Fe_{19}Ni_{81}$ alloy (permalloy, Py). The whole film structures are represented by $TiO_x$(2)/Py(3)/$WO_2$(20), where the number in parenthesis indicates the thickness in nanometer [Fig. 3(a)]. After the $WO_2$ growth with the sputtering parameters as stated above, Ti and Py were deposited *in situ* by a radio-frequency magnetron sputtering at Ar deposition pressure of 0.4 Pa. The $TiO_x$ layer is obtained from the as-deposited Ti metal by natural oxidation in the air. The purpose of the $TiO_x$ layer is to prevent the Py layer from being oxidized; we can neglect current shunting by the $TiO_x$ layer due to its high resistance. We also prepared a Py(3)/W(4) bilayer as a control sample. The Py, W, and $TiO_x$ thickness were estimated from growth rate of each layer determined by XRR measurement beforehand. The resistivity of W and Py is 180 and 100 μΩcm, respectively, obtained from the relation between the inverse sheet resistance and the Py thickness (not shown); a good linear relation is found when the Py thickness ranges from 3 to 6 nm, indicating that an increase of the resistivity stemming from interface effects is negligibly small. While the W thickness of 4 nm is chosen by reference to the previous studies[1,2,14], we set the $WO_2$ thickness to 20 nm so that the resistance of the $WO_2$ layer matches with that of the W layer; the current and the SOT data quality is expected to be roughly comparable between the W and $WO_2$ layers. The Py/$WO_2$ (Py/W) bilayer was fabricated into a Hall bar with two arms by photolithography and Ar ion milling. Ti(5)/Pt(60) contact pads were attached at the end of devices for electrical measurement. The Hall bar has channel dimensions of 50 μm length and 10 μm width, as shown in Fig. 3(a). The $\phi$ represents the azimuthal angles of the external magnetic field ($B_{ext}$). We apply an ac current $I_{ac} = \sqrt{2}I_{rms}\sin(2\pi f t)$ with root mean square of current $I_{rms}$ in the *x* axis direction and frequency $f = 13$ Hz. We define the current flow direction (*x* axis) as the $(10\bar{1}0)$-orientation of the $Al_2O_3$ (0001) substrate; note that this choice has no significance because we cannot expect anisotropy in any transport measurements considering that the $WO_2$ film consists of the multidomains discussed in Fig. 1(d). We set $I_{rms}$ to be 0.2 mA for longitudinal resistance ($R$) measurement in *x*-axis direction, and 0.2 mA for Hall resistance ($R_H$) measurement and 2.0 mA for harmonic Hall measurement in the *y*-axis directions.

We evaluate the SOT by conducting harmonic Hall measurement[4,5,12]. By applying the $I_{ac}$ to the Hall bar, the SOT is induced at FM/NM interface, which has two components with different symmetries, namely, DL and FL SOTs[4,12,37]. These SOTs correspond to DL effective field $B_{DL} \parallel (\sigma \times m)$ and FL effective field $B_{FL} \parallel \sigma$, where $m$ and $\sigma$ is the direction of the magnetization and accumulated spin polarization; the latter is along the $y$-axis direction. Of these two SOTs, the DL SOT is most relevant to the magnetization switching[4,37]. In order to distinguish the $B_{DL}$ and $B_{FL}$, the harmonic Hall measurements were performed by measuring an angle dependence of first and second harmonic resistance ($R_H^{1\omega}$, $R_H^{2\omega}$) at fixed $B_{ext}$ in $xy$ plane; the applied $B_{ext}$ varies 0.1 to 1.2 T. The $R_H{}^{1\omega}$ corresponds to the conventional dc Hall measurement, while the $R_H^{2\omega}$ reflects the influence of SOT; the former and the latter obey following Eq. (1) and Eq. (2)[5].

$$R_H^{1\omega} = R_{PHE} \sin 2\phi \sin^2\theta \quad (1)$$

$$R_H^{2\omega} = -\left(R_{AHE}\frac{B_{DL}}{B_{ext}+B_k}+R_{\nabla T}\right)\cos\phi + 2R_{PHE}\frac{B_{FL}+B_{Oe}}{B_{ext}}\left(2\cos^3\phi - \cos\phi\right) \quad (2)$$

$$\equiv -R_{DL+\nabla T}\cos\phi + R_{FL+Oe}\left(2\cos^3\phi - \cos\phi\right) \quad (3)$$

Here, $R_{PHE}$, $R_{AHE}$, $B_k$, $R_{\nabla T}$, and $B_{Oe}$ correspond to planar Hall resistance, anomalous Hall resistance, out-of-plane anisotropy field, thermal induced second-harmonic resistance driven by a temperature gradient, *e.g.* the anomalous Nernst effect[38] and spin Seebeck effect[39], and current induced Oersted field, respectively. $R_{AHE}$ and $B_k$ were estimated by Hall resistance depending on the out-of-plane $B_{ext}$. The $\phi$ dependence of $R_H^{2\omega}$ is induced by the small modulation of the magnetization from its equilibrium position due to the current-driven SOTs. We can define the $B_{DL}$ contribution ($R_{DL+\nabla T}$) and the $B_{FL}$ contribution ($R_{FL+Oe}$) as the coefficients of the $\cos\phi$ and ($2\cos^3\phi - \cos\phi$) in Eq. (3). Figure 3(b) shows $R_H^{1\omega}$ as a function of $\phi$ with $B_{ext}$ = 0.5 T; $R_{PHE}$ = 0.42 Ω was obtained as the fitting results in accordance with Eq. (1). The top panel of Figure 3(c) shows the corresponding $R_H^{2\omega}$ measured at 0.5 T, which is well fitted by Eq. (2). We estimated the $R_{DL+\nabla T}$ and $R_{FL+Oe}$ from the $R_H^{2\omega}$ fitting, which is shown in the bottom panel in Fig. 3(c), respectively; the result indicates the large $R_{DL+\nabla T}$ contribution and the small $R_{FL+Oe}$ contribution to the $R_H^{2\omega}$ in Py/$WO_2$ bilayer.

Then we extract two SOT effective fields $B_{DL}$ and $B_{FL}$ through the coefficients $R_{DL+\nabla T}$ and $R_{FL+Oe}$. Figure 4(a) shows the $R_{DL+\nabla T}$ for the Py/$WO_2$ bilayer and Py(3)/W(4) as control sample. The data indicates linear dependence on $1/(B_{ext} + B_k)$ in accordance with Eq. (2). The data point in the low-field region for Py/W bilayer is deviated from the linear fitting, possibly due to unsaturated in-plane magnetization[23]. The slopes and intercepts of the $R_{DL+\nabla T}$ correspond to the $B_{DL}$ and the $R_{\nabla T}$, respectively. We estimated the DL effective field per current density $B_{DL}/J$, where $J$ is the applied charge current density flowing in the $WO_2$ (W) layer. The estimated $B_{DL}/J$ is +3.03 mT/($10^{11}$Am$^{-2}$) for Py/$WO_2$ and −4.58 mT/($10^{11}$Am$^{-2}$) for Py/W, suggesting the sizable DL SOT generation in Py/$WO_2$ as well as in Py/W. The $R_{\nabla T}$ of Py/$WO_2$ is much larger than Py/W.

The temperature gradient in Py/$WO_2$ is larger compared with Py/W due to the higher resistivity of $WO_2$ than that of W, resulting in the large $R_{\nabla T}$. The $R_{FL+Oe}$ for Py/$WO_2$ and Py/W bilayers are plotted as a function of $1/B_{ext}$ as shown in Fig. 4(b), which is well explained by Eq. (2). The $B_{FL}$ were obtained by subtracting the $B_{Oe}$ contribution estimated by Ampere's law as $B_{Oe} = \mu_0 Jd/2$, where $\mu_0$ is the magnetic permeability in a vacuum, and $d$ is the thickness of $WO_2$ (W). The $B_{FL}/J$ for Py/$WO_2$ and Py/W are $+0.75 \times 10^{-1}$ and $-0.30$ mT/($10^{11}$Am$^{-2}$), respectively.

In order to further discuss the SOT, we evaluate the efficiency $\xi_{DL(FL)}$ from the $B_{DL(FL)}/J$ using following equation[40]:

$$\xi_{DL(FL)} = \frac{2e\mu_0 M_s t_{Py}}{\hbar} \frac{B_{DL(FL)}}{J} \tag{4}$$

where $e$, $M_s$, and $\hbar$ are the elementary charge, the saturation magnetization, and the Dirac constant, respectively. The $M_s$ for Py/$WO_2$ and Py/W are $6.2 \times 10^5$ and $6.7 \times 10^5$ A/m, respectively, measured by the superconducting quantum interference device magnetometer, in agreement with those for typical Py thin film[6,8,10,16,17,22]. The $\xi_{DL}$ and $\xi_{FL}$ for $WO_2$ and W are summarized in Fig. 4(c). We obtain $\xi_{DL} = +0.174 \pm 0.021$ for $WO_2$ and $-0.281 \pm 0.006$ for W. The $\xi_{DL}$ for W is quantitatively consistent with the previous reports including its negative sign[14–16], demonstrating validity of our experimental setup. Note that while recent reports point out the importance of the DL-SOT generation from single FM layers[41-43], we experimentally confirmed that the DL-SOT efficiency of single 3-nm-thick Py layer is as small as 0.001. In our bilayers, the contributions from the Py layer is thus negligible compared to the above-mentioned DL-SOT efficiencies of $WO_2$ and W. We also consider that the $\xi_{DL}$ for $WO_2$ is saturated since the $WO_2$ layer thickness of 20 nm is sufficiently above the spin-diffusion length. The prior studies have shown that the typical spin-diffusion length is ~2 nm for non-magnets with efficient spin current generation such as Pt, W, and Ir oxide[4-10,16,22], suggesting that the spin diffusion length of $WO_2$ is also around 2 nm; this rough estimate would not affect the above-mentioned $\xi_{DL}$ as far as the $WO_2$ thickness is much longer than the spin diffusion length. In terms of magnitude, the $\xi_{DL}$ of +0.174 for the $WO_2$ is about two thirds of that for the W, reinforcing the validity of $WO_2$ as the spintronic material; the $\xi_{DL}$ is indeed comparable to those for Pt[4,6,22,23], Ta[5,11,13], and amorphous $IrO_2$[22,23]. Regarding the FL-SOT, we obtain $\xi_{FL} = +0.042 \pm 0.003$ for $WO_2$ and $-0.018 \pm 0.002$ for W. The smaller $\xi_{FL}$ with the same sign compared to their $\xi_{DL}$ is likely to stem from the bulk SHE; the similar trend on the magnitude and sign is found in typical 5$d$ TMs[4–6,13,16,23] as well. The larger $\xi_{DL}$ than $\xi_{FL}$ is helpful for demonstrating the magnetization switching since the DL-SOT plays an important role in magnetization control.

We then focus on the observed positive sign of $\xi_{DL}$ for $WO_2$. This contrasts with the negative sign of $\xi_{DL}$ for W, reflecting the difference of the electronic structure between $WO_2$ and W. The 5$d$ electron solely dominates the density of states at $E_F$ in $WO_2$ and hence the spintronic behavior of $WO_2$ can be totally different with W and/or slightly oxidized W. For 4$d$ and 5$d$ TMs, the theoretical study concludes that the spin Hall conductivity is proportional to the SOC constant ($\lambda$)[44]; the spin Hall conductivities of TMs show a change from the negative to positive signs with increasing electron numbers since the sign of the $\lambda$ is reversed between more-than half and less-than half filling in 5*d* orbitals; this sign change is experimentally confimerd[45]. In contrast, the spin Hall conductivity of 4*d* TMOs is theoretically predicted to be dominated by the quadratic term of the $\lambda$[46], conveying that the 4*d* TMOs keeps the same sign irrespective of the sign of the $\lambda$ which depends on the 5*d* electron number. Since the positive sign is observed in the rutile $IrO_2$[21], the positive sign of the $WO_2$ is plausible by assuming that the quadratic term of the $\lambda$ is dominant in the 5*d* oxides as theoretically predicted in 4*d* oxides. Thus, the $WO_2$ would be an attractive spintronic material, motivating us to gain further insight from both experimental and theoretical aspects. For instance, the anisotropic effect on the SOT may bring rich information about the relationship between the electronic structure and the spin-current properties since the anisotropic electronic structure is expected in the rutile compounds. At current stage, while we cannot pin down the positive sign for $WO_2$ in detail, our finding is a good starting point to clarify the spin current physics behind the unique electronic structure inherent to 5*d* TMOs.

In conclusion, we studied an efficient room temperature generation of spin current in Py/$WO_2$ bilayer, where the distorted rutile type $WO_2$ is epitaxially grown on $Al_2O_3$ substrate by reactive sputtering method. By tuning the sputtering power and oxygen ratio, we obtained the monoclinic $WO_2$ (010) characterized by the XRD. Further film characterization revealed the metallic character and the $5d^2$ electron configuration of $WO_2$, supported by the resistivity and XPS measurements. By performing the harmonic Hall measurement, we estimated that the $\xi_{DL}$ (+0.174) of the $WO_2$ is roughly two thirds of that (–0.281) of W control sample in terms of magnitude and is as large as those values in typical Pt and Ta, and $IrO_2$, demonstrating efficient spin current generation from the $WO_2$. The positive sign of $\xi_{DL}$ for $WO_2$ in contrast to the negative one for W, is the same with that for $IrO_2$, which is consistent with theoretical prediction based on the 4*d* oxides. Our findings indicate that $WO_2$ is a promising spintronic material, which may be a component of magnetic memory device driven by SOT switching.

## ACKNOWLEDGEMENT

The authors thank H. Kontani for fruitful discussion, K. Omura for the XPS measurement, and T. Arakawa for technical support. This work was carried out at the Center for Advanced High Magnetic Field Science in Osaka University under the Visiting Researcher's Program of the Institute for Solid State Physics, the University of Tokyo. This work was supported by Nanotechnology Platform of MEXT, Grant Number JPMXP09S21OS0027, the JSPS KAKENHI (Grant Nos. JP19K15434, JP19H05823, and 22H04478), JPMJCR1901 (JST-CREST), the Spintronics Research Network of Japan (Spin-RNJ), Nippon Sheet Glass foundation for Materials Science and Engineering, and Iketani Science and Technology Foundation. We acknowledge stimulating discussions at the meeting of the Cooperative Research Project of the Research Institute of Electrical Communication, Tohoku University

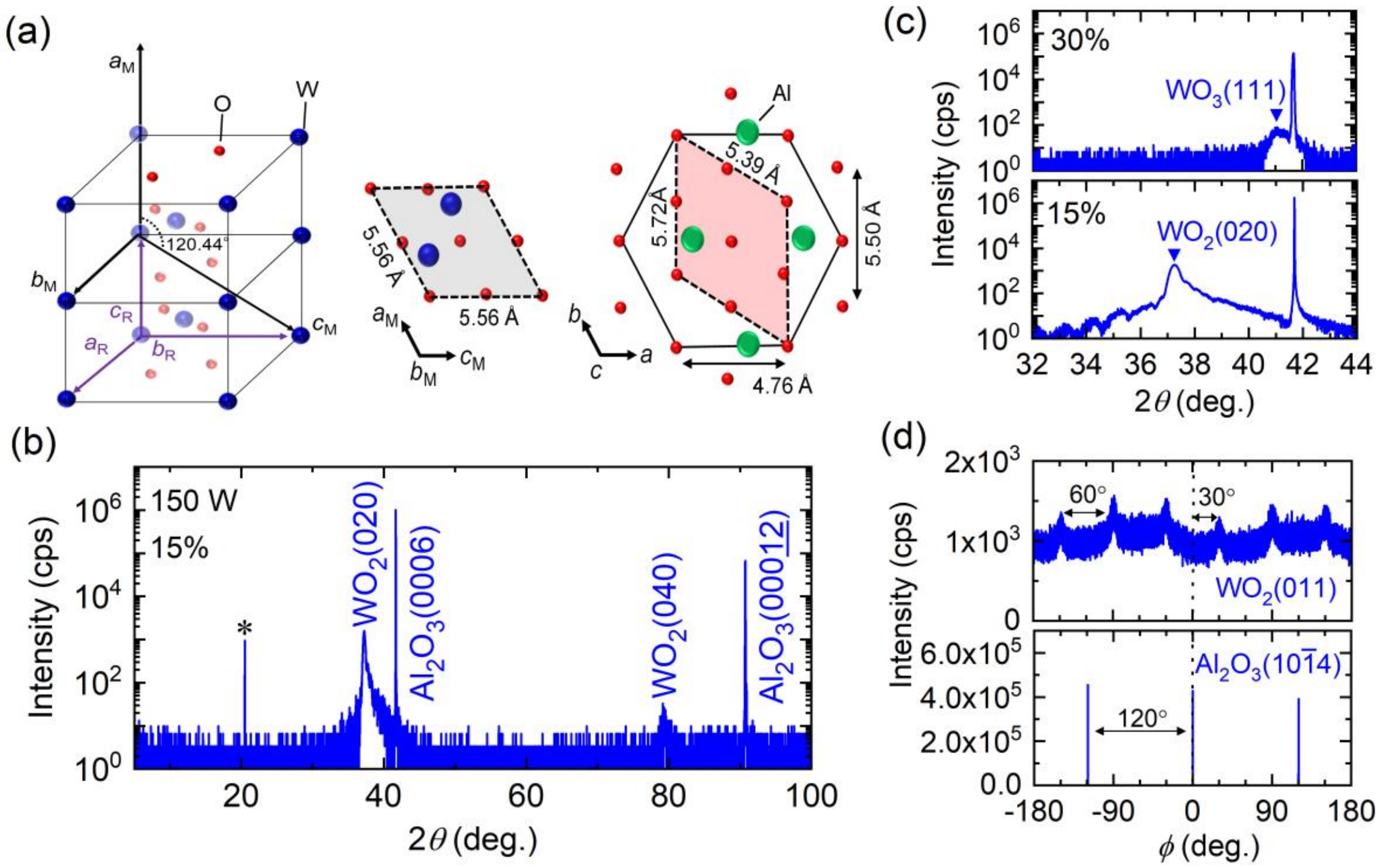

**Fig. 1.** (a) (left) Illustration of structure for monoclinic (*M*) $WO_2$ (space group #14 $P2_1/c$) and pseudrutile (*R*) $WO_2$. (right) The crystallographic relationship between the monoclinic $WO_2(010)$ and the $Al_2O_3(0001)$. (b) X-ray diffraction (XRD) $2\theta$-$\theta$ scan of a $WO_2$ film grown on the $Al_2O_3(0001)$ substrate at sputtering condition with partial oxygen of 15% and the power of 150 W. The asterisk indicates the forbidden (0003) reflections of $Al_2O_3$. (c) Magnified view of the scan around $WO_2(020)$ reflection in (b) and the XRD pattern around pseudcubic $WO_3(111)$ reflection at the condition with partial oxygen of 30%. (d) The azimuthal $\phi$ scan of $(10\bar{1}4)$ diffraction for $Al_2O_3$ substrate (bottom) and of (011) diffraction for $WO_2$ film (top).

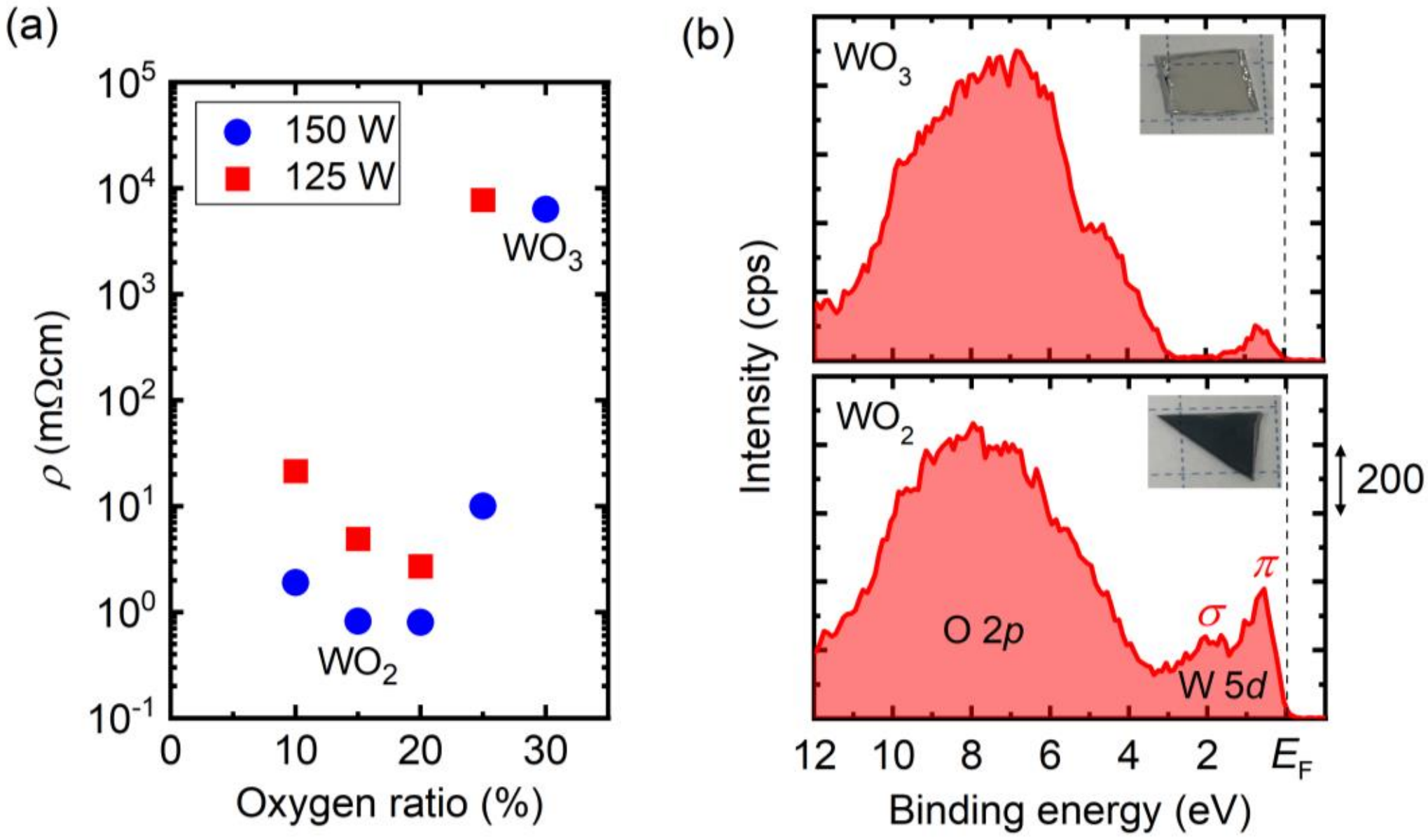

**Fig. 2.** (a) Electrical resistivity as a function of the partial oxygen ratio with respect to working pressure of 0.4 Pa at sputtering power of 125 W and 150 W. (b) Valence band spectra measured by *x*-ray photoelectron spectroscopy for the $WO_2$ and $WO_3$ films, respectively. Insets are photographs of the films.

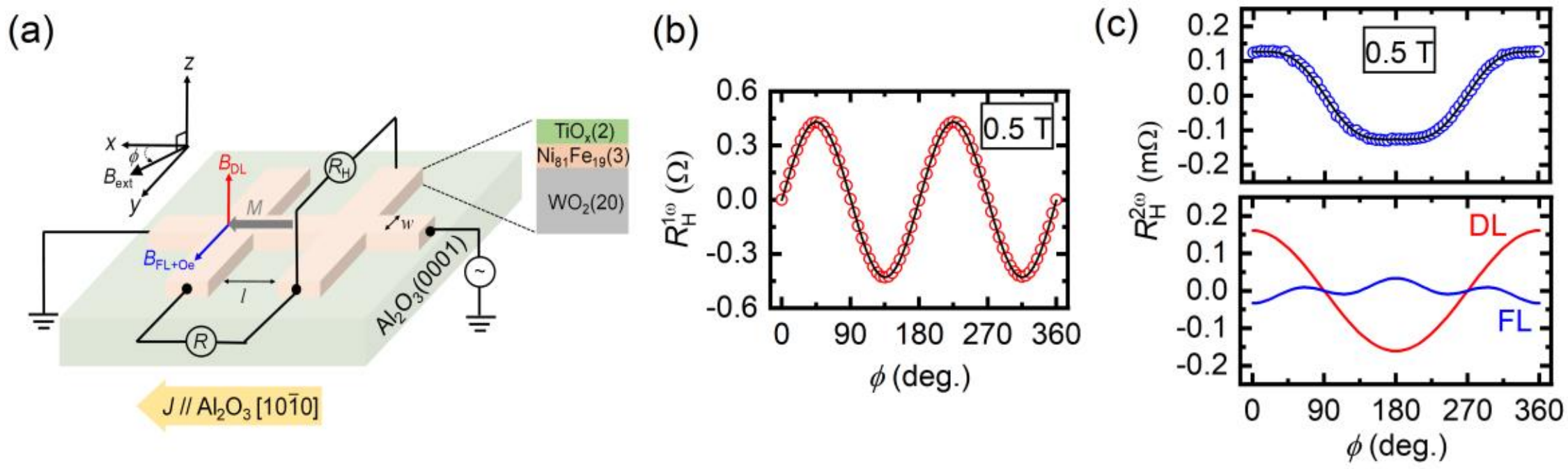

**Fig. 3.** (a) Schematic illustration of the device and cross section of the samples. The $\phi$ represents the azimuthal angles of the external magnetic field ($B_{ext}$). AC current is applied along the $x$ axis direction to detect Hall resistance ($R_H$) in the $y$-axis direction, and detect longitudinal resistance ($R$) in the $x$-axis direction. (b) First-harmonic Hall resistance ($R_H^{1\omega}$) of Py(3)/$WO_2$(20) measured at 0.5 T. The solid curve is fit to the data using Eq. (1). (c) The corresponding second-harmonic Hall resistance ($R_H^{2\omega}$) with curve fit using Eq. (2) (top). The data is separated to cos$\phi$ and (2cos$^3\phi$–cos$\phi$) components (bottom) from the fitting result, which indicate DL contribution and FL contribution, respectively.

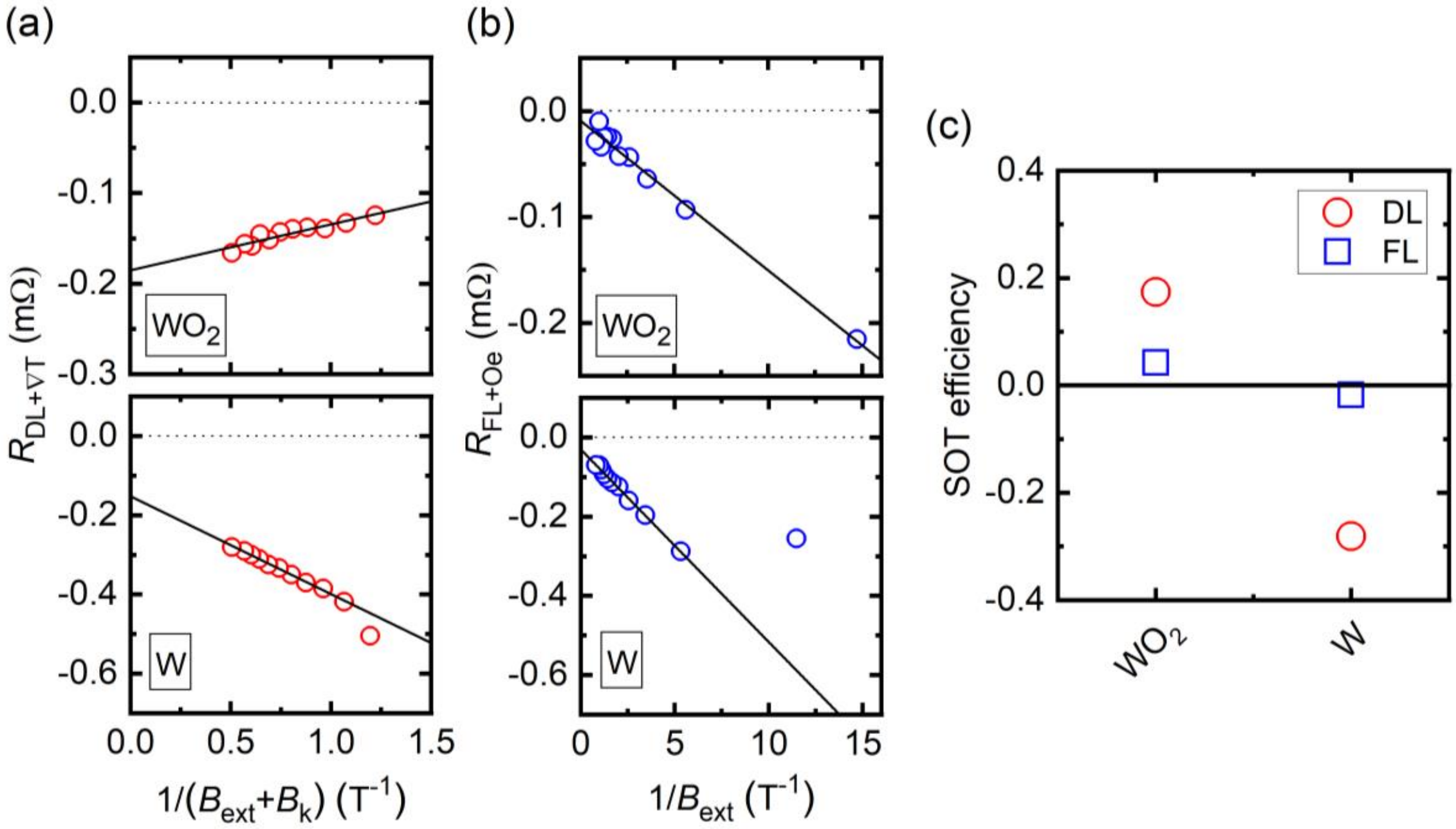

**Fig. 4.** (a) $R_{DL+\nabla T}$ as a function of 1/($B_{ext}$ + $B_k$) in Py(3)/$WO_2$(20) and Py(3)/W(4) samples. (b) $R_{FL+Oe}$ as a function of 1/$B_{ext}$ in Py(3)/$WO_2$(20) and Py(3)/W(4) samples. The solid lines are linear fits to the experimental data. (c) Dampinglike and fieldlike SOT efficiencies for $WO_2$ and W. The error bars are smaller than the symbol size.